\makeatletter
\declare@file@substitution{revtex4-1.cls}{revtex4-2.cls}
\makeatother

\documentclass[twocolumn,trackchanges]{aastex631}

\usepackage{graphicx}
\usepackage{amsmath}
\usepackage{url}
\usepackage{enumerate}
\usepackage{enumitem}
\usepackage{multirow}
\usepackage{soul}
\usepackage{bm}
\usepackage{float}

\hypersetup{
	colorlinks=true,
	linkcolor=green,
	anchorcolor=red,
	citecolor=blue,
	filecolor=red,
	pagecolor=red,
	urlcolor=red
}
\usepackage{url}
\usepackage{soul}
\usepackage{colortbl}

\RequirePackage{color}

\newfont{\myfont}{cmmib10}

\def\lapprox{\mathrel{\hbox{\rlap{\hbox{\lower4pt\hbox{$\sim$}}}\hbox{$<$}}}}
\def\gapprox{\mathrel{\hbox{\rlap{\hbox{\lower3pt\hbox{$\sim$}}}\hbox{$>$}}}}

\definecolor{apricot}{rgb}{0.88,0.81,0.65}

\graphicspath{{./}{figures/}}

\begin{document}

	\title{Investigation of profile shifting and subpulse movement in PSR J0344$-$0901 with FAST}

	\correspondingauthor{R. Yuen, N. Wang}
	\email{ryuen@xao.ac.cn}
	\email{na.wang@xao.ac.cn}
	
	\author{H. M. Tedila}
	
	\affiliation{National Astronomical Observatories, Chinese Academy of Sciences, A20 Datun Road, Chaoyang District, Beijing 100101, China}
	\affiliation{Key Laboratory of Radio Astronomy and Technolgoy, Chinese Academy of Sciences, A20 Datun Road, Chaoyang District, Beijing, 100101, P.R. China}
	\affiliation{Xinjiang Astronomical Observatory, Chinese Academy of Sciences, 150 Science 1-Street, Urumqi, Xinjiang, 830011, People's Republic of China}
	\affiliation{University of Chinese Academy of Sciences, 19A Yuquan Road, 100049 Beijing, People's Republic of China}
	\affiliation{Arba Minch University, Arba Minch 21, Ethiopia}
	
	\author{R. Yuen}
	\affiliation{Xinjiang Astronomical Observatory, Chinese Academy of Sciences, 150 Science 1-Street, Urumqi, Xinjiang, 830011, People's Republic of China}
	\affiliation{Xinjiang Key Laboratory of Radio Astrophysics, 150 Science1-Street, Urumqi, Xinjiang, 830011, People's Republic of China}
	
	\author{N. Wang}
	\affiliation{Xinjiang Astronomical Observatory, Chinese Academy of Sciences, 150 Science 1-Street, Urumqi, Xinjiang, 830011, People's Republic of China}
	\affiliation{Xinjiang Key Laboratory of Radio Astrophysics, 150 Science1-Street, Urumqi, Xinjiang, 830011, People's Republic of China}
	
	\author{D. Li}
	\affiliation{National Astronomical Observatories, Chinese Academy of Sciences, A20 Datun Road, Chaoyang District, Beijing 100101, China}
	\affiliation{NAOC-UKZN Computational Astrophysics Centre, University of KwaZulu-Natal, Durban 4000, South Africa}
	\affiliation{CAS Key Laboratory of FAST, NAOC, Chinese Academy of Sciences, Beijing 100101, People's Republic of China}
	\affiliation{Key Laboratory of Radio Astronomy and Technolgoy, Chinese Academy of Sciences, A20 Datun Road, Chaoyang District, Beijing, 100101, P.R. China}
	\affiliation{University of Chinese Academy of Sciences, 19A Yuquan Road, 100049 Beijing, People's Republic of China}
	
	\author{Z. G. Wen}
	\affiliation{Xinjiang Astronomical Observatory, Chinese Academy of Sciences, 150 Science 1-Street, Urumqi, Xinjiang, 830011, People's Republic of China}
	\affiliation{Xinjiang Key Laboratory of Radio Astrophysics, 150 Science1-Street, Urumqi, Xinjiang, 830011, People's Republic of China}
	
	\author{W. M. Yan}
	\affiliation{Xinjiang Astronomical Observatory, Chinese Academy of Sciences, 150 Science 1-Street, Urumqi, Xinjiang, 830011, People's Republic of China}
	\affiliation{Xinjiang Key Laboratory of Radio Astrophysics, 150 Science1-Street, Urumqi, Xinjiang, 830011, People's Republic of China}
	
	\author{J. P. Yuan}
	\affiliation{Xinjiang Astronomical Observatory, Chinese Academy of Sciences, 150 Science 1-Street, Urumqi, Xinjiang, 830011, People's Republic of China}
	\affiliation{Xinjiang Key Laboratory of Radio Astrophysics, 150 Science1-Street, Urumqi, Xinjiang, 830011, People's Republic of China}
	
	\author{X. H. Han}
	\affiliation{Xinjiang Astronomical Observatory, Chinese Academy of Sciences, 150 Science 1-Street, Urumqi, Xinjiang, 830011, People's Republic of China}
	
	\author{P. Wang}
	\affiliation{National Astronomical Observatories, Chinese Academy of Sciences, A20 Datun Road, Chaoyang District, Beijing 100101, China}
	\affiliation{Key Laboratory of Radio Astronomy and Technolgoy, Chinese Academy of Sciences, A20 Datun Road, Chaoyang District, Beijing, 100101, P.R. China}
	
	\author{W. W. Zhu}
	\affiliation{National Astronomical Observatories, Chinese Academy of Sciences, A20 Datun Road, Chaoyang District, Beijing 100101, China}
	\affiliation{Key Laboratory of Radio Astronomy and Technolgoy, Chinese Academy of Sciences, A20 Datun Road, Chaoyang District, Beijing, 100101, P.R. China}
	
	\author{S. J. Dang}
	\affiliation{School of Physics and Electronic Science, Guizhou Normal University, Guiyang, 550001, People’s Republic of China}
	
	\author{S. Q. Wang}
	\affiliation{Xinjiang Astronomical Observatory, Chinese Academy of Sciences, 150 Science 1-Street, Urumqi, Xinjiang, 830011, People's Republic of China}
	
	\author{J. T. Xie}
	\affiliation{Research Center for Intelligent Computing Platforms, Zhejiang Laboratory, Hangzhou, 311100, China}
	
	\author{Q. D. Wu}
	\affiliation{Xinjiang Astronomical Observatory, Chinese Academy of Sciences, 150 Science 1-Street, Urumqi, Xinjiang, 830011, People's Republic of China}
	\affiliation{University of Chinese Academy of Sciences, 19A Yuquan Road, 100049 Beijing, People's Republic of China}

	\author{Sh. Khasanov}
	\affiliation{Institute of Modern Physics, Chinese Academy of Sciences, Lanzhou, 730000, China}
	\affiliation{University of Chinese Academy of Sciences, 19A Yuquan Road, 100049 Beijing, People's Republic of China}
	\affiliation{Samarkand State University, Samarkand 140104, Uzbekistan}
	
	\author{FAST Collaboration}

	\begin{abstract}
		
		We report two phenomena detected in PSR J0344$-$0901 from two observations conducted at frequency centered at 1.25 GHz using the Five-hundred-meter Aperture Spherical radio Telescope (FAST). The first phenomenon manifests as shifting in the pulse emission to later longitudinal phases and then gradually returns to its original location. The event lasts for about 216 pulse periods, with an average shift of about $0.7^\circ$ measured at the peak of the integrated profile. Changes in the polarization position angle (PPA) are detected around the trailing edge of the profile, together with an increase in the profile width. The second phenomenon is characterized by the apparent movement of subpulses, which results in different subpulse track patterns across the profile window. For the first time in this pulsar, we identify four emission modes, each with unique subpulse movement, and determine the pattern periods for three of the emission modes. Pulse nulling was not detected. Modeling of the changes in the PPA using the rotating vector model gives an inclination angle of $75.12^\circ \pm 3.80^\circ$ and an impact parameter of $-3.17^\circ \pm 5.32^\circ$ for this pulsar. We speculate that the subpulse movement may be related to the shifting of the pulse emission.

	\end{abstract}


	\section{Introduction}
	\label{sect:intro}
	
	A large diversity of emission variation at different timescales has been observed in radio pulsars. This includes the familiar examples of subpulse drifting, pulse nulling, and profile mode-changing. More than half of the radio pulsars sampled \citep{Weltevrede06, Weltevrede07} was found with measurable drift of subpulses from one pulse to the next through the integrated profile window. There is also evidence from both radio and $\gamma$-ray observations that a pulsar magnetosphere may exhibit some kind of changes, which affect the subpulse drift pattern \citep{SLS00, LHK+10, Kramer2006, ABB+13, KSJ13}. From the observations of pulsars with multiple subpulse drift modes \citep{SMK05}, each with particular drift rate and pattern, the subpulse drift rate can vary abruptly and then returns to its initial value. In addition, the changes between different subpulse drift modes have been observed to correlate with changes in the integrated profile \citep{WF81, RWR05, Joshi12, RWB13}. However, changes may also be gradual. The observation of PSR B0809+74  reveals that a change in the subpulse drift rate following a null exhibits a slow adjustment back to its normal drift rate over tens of pulse periods. 
	
	On longer timescales, many individual pulses from a pulsar are averaged to form the integrated profile whose shape is stable over a long time \citep{HMT75, Pulsars1977, RR95}. Sudden changes in the profile shape, known as profile mode-changing, have also been detected in many pulsars. However, variations that occur on individual pulse level or over several tens of pulses are hidden by this averaging method. One such variation is the observed shifting of the radio profile in PSR J0344$-$0901. During the event, the peak of the profile, obtained from averaging the associated pulses, exhibits a shift towards later longitudinal phases to a new location, where the profile remains for tens of pulses, and then returns to its original position. In addition, the profile intensity drops, and the profile shape changes. The pulsar was first reported by \citet{CLH+20} for its timing properties, but without giving detailed description on the shifting phenomenon. Radio profile shifts have been detected in a few pulsars \citep{RRW06, Perera15, HHP+16,WOR+16, Shaifullah18}, where the observed shifts are all toward earlier longitudinal phases, as opposed to that in PSR J0344$-$0901. Profile shifting was suggested in association with relativistic aberration and retardation effects in conventional geometric models \citep{CRZ00, DK04, Gangadhara05}. However, the effects are long-lasting in contrast to the sporadic changes observed in PSR J0344$-$0901.
	
	In this paper, we present the emission variation observed in PSR J0344$-$0901. The pulsar was discovered in September 2017 during the pilot scans of the Commensal Radio Astronomy FAST Survey \citep[CRAFTS;][]{Li18} using the ultra-wide-bandwidth (UWB) receiver with the Parkes Digital Filter Bank backend system (PDFB4), and confirmed by the Parkes 64-m Radio Telescope \citep{CLH+20}. It has a pulse period of \( P = 1.23 \, \text{s} \), and a period derivative of \( \dot{P} = 3.4708 \times 10^{-15} \, \text{s} \, \text{s}^{-1} \). In addition, the pulsar possesses a surface dipolar magnetic field of \( B_{\rm d} = 2.06 \times 10^{12} \, \text{G} \), a characteristic age of 5.58 Myr, and a spin-down energy loss of \( \dot{E} = 7.44 \times 10^{31} \, \text{erg} \, \text{s}^{-1} \). The dispersion measure (DM) is given by \( 31 \, \text{cm}^{-3} \, \text{pc} \) \citep{CLH+20}. In Section \ref{sect:analysis}, we outline the observations and the procedures for data analysis. The classification of different emission pulses is defined in Section \ref{sect:emission_classify}. Analysis of the pulse sequences in relation to profile shifting is presented in Section \ref{sect:Analysis_sequences}. The examination of the subpulse movement is given in Section \ref{sect:drifting}. We briefly discuss the implications of our results and the main emission features of the pulsar in Section \ref{sect:discussion}, and summarize our results in Section \ref{sect:summary}.
	
	\begin{figure*}
		\centering  
		\includegraphics[width=2\columnwidth]{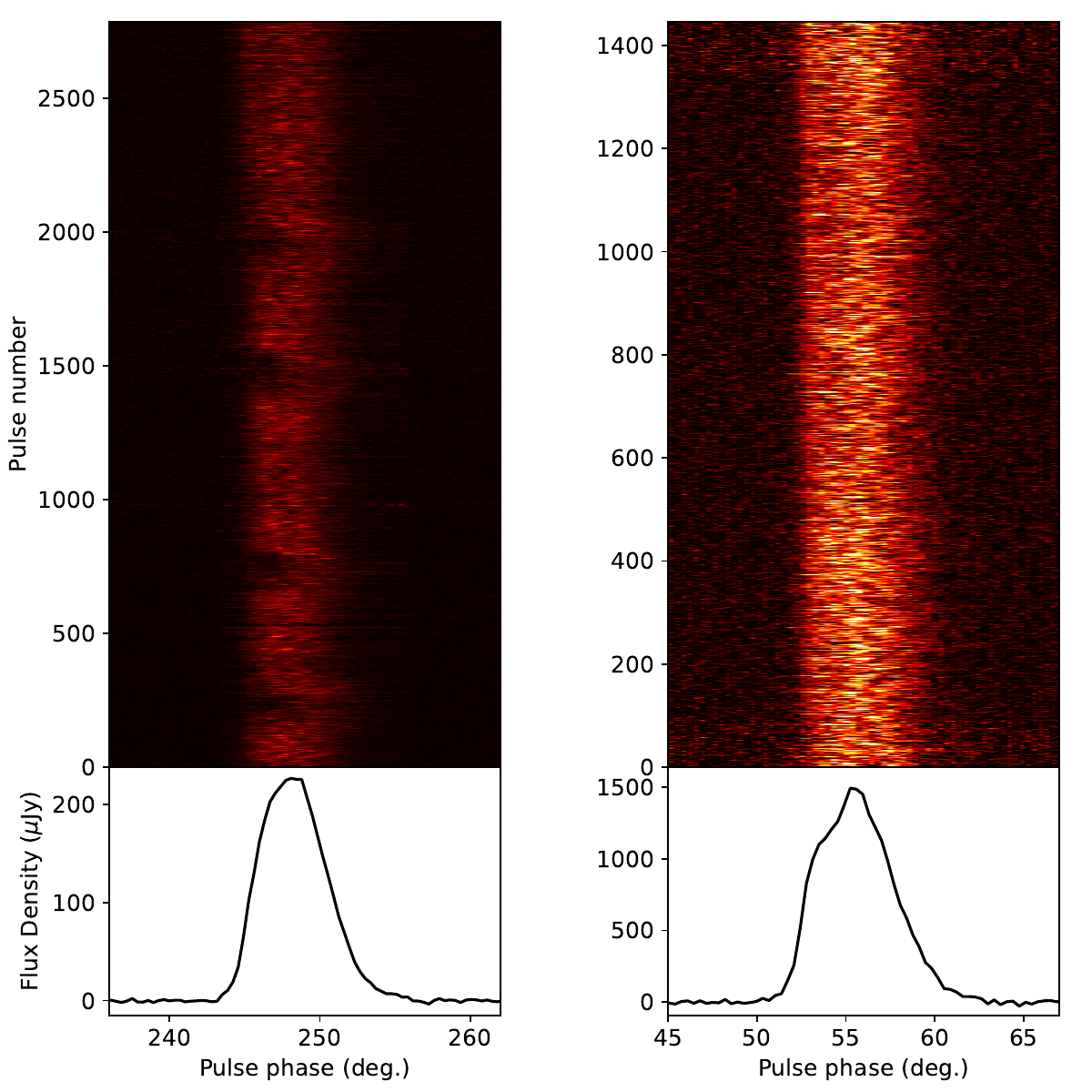}
		\caption{The sequences of single pulses from Observations I (left) and II (right) are shown, and the integrated profile for each is presented in the lower panel.}
		\label{pulse_sequences}
	\end{figure*}
	
	\section{The observations and data analysis}
	\label{sect:analysis}

	The observations of PSR J0344$-$0901 were conducted using the FAST telescope with the 19-beam receiver and the Reconfigurable Open Architecture Computing Hardware--version 2 (ROACH2) signal processor \citep{Jiang19, Jiang20}. The observations spanned a frequency range from 1.05 to 1.45 GHz. For Observation I, the time and frequency resolutions were set at 49.152 µs and 0.122 MHz, respectively. Observation II was conducted with time and frequency resolutions of 49.152 µs and 0.488 MHz, respectively. The data was recorded in the search-mode PSRFITS format \citep{Hotan04}, incorporating four polarizations. Observation I employed 4096 frequency channels, while Observation II used 1024 frequency channels each. More information for the observations is provided in Table \ref{table:observations}. In addition, flux calibration for Observation II could not be performed due to absence of the off-track data.
	
	\begin{table}
		\setlength{\tabcolsep}{3.2 pt}
		\centering
		\caption{The two single-pulse observations for PSR J0344$-$0901. The total duration (in seconds), the total number of pulses, the number of channels ($n_{\rm chan}$), and  the frequency resolutions (chbw) in MHz used for each observation are shown in columns 3, 4, 5, and 6, respectively.}
		\begin{tabular}{c|cccccc} 
			\hline
			Observation & Date & Duration & Pulses & $n_{\rm chan}$ &chbw \\
			& & (s) &  & &(MHz)\\
			\hline
			I & 2019-04-23 & 3450 & 2815 & 4096&0.122\\
			II & 2020-09-19 & 1800 & 1466 & 1024&0.488\\
			\hline
		\end{tabular}
		\label{table:observations}
	\end{table}

	\begin{figure*}
		\centering
		\includegraphics[width=1.85\columnwidth]{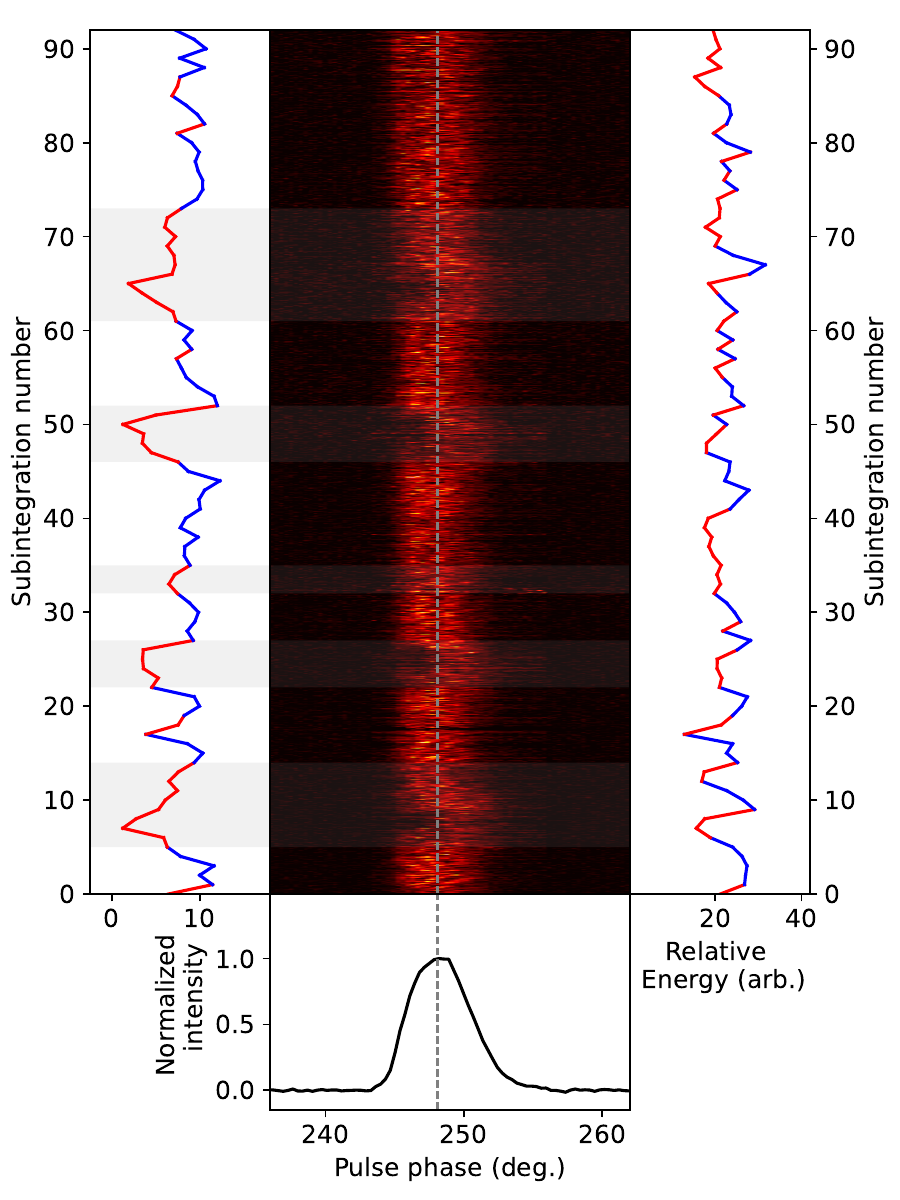}
		\caption{Plot in the upper middle panel shows the pulse sequence from Observation I with sub-integrations of every 30 pulses. The pulse energy variation in the leading part of the profile is demonstrated in the upper left plot, where the energy above and below the threshold is signified by blue and red, respectively. The shifting events are indicated by regions in gray, and an accompanying drop in intensity is also apparent in the leading component of the profile. In the upper right plot, the energy variation in the trailing part of the profile is shown. The integrated pulse profile obtained from all pulses in Observation I is given in the lower plot, and the vertical dashed line in gray demarcates the range for the leading and trailing parts of the profile.}
		\label{fig-Energy_criteria}
	\end{figure*}
	
	The first step in data analysis involved the use of the {\tt\string DSPSR} software package \citep{Straten11} to de-disperse and produce the single-pulse integrations. Spectral band edges were zapped using {\tt\string paz}, and the radio frequency interference (RFI) was removed from the data using {\tt\string pazi} in the {\tt\string PSRCHIVE} software package \citep{Hotan04}. The data was then calibrated using the {\tt\string PAC} program in the {\tt\string PSRCHIVE} software package to flatten the bandpass and transform the polarization products to Stokes parameters. After that, the frequency channels and polarization of the data were compressed. The radiometer equation, 
	\begin{equation}\label{radiometer}
		S_{\rm av} = \frac{S_{\rm sys}}{\sqrt{n_{\rm p}f_{\rm b} t_{\rm s}}},
	\end{equation}
	was then used to calibrate the observed pulses based on the off-pulse noise \citep{Dicke1982, wen21}. Here, $n_{\rm p}=4$, $f_{\rm b} = 400\,{\rm MHz}$ and $t_{\rm s} = 49.152\,\mu{\rm s}$ are the number of polarization, effective frequency bandwidth and the sampling time, respectively, and $S_{\rm sys}\approx 1\,{\rm Jy}$ is the system equivalent flux density (SEFD) of FAST \citep{Yu17}. Flux calibration was then performed on all pulse profiles by multiplying with $S_{\rm av}$, and the estimated flux density ($S_{\rm av}$) for this pulsar is about 3.5\,{\rm mJy} by averaging the fluxes in all pulse phases and time.

	\section{Different pulse emission} 
	\label{sect:emission_classify}
	
	The pulse sequences of our observations demonstrate significant emission fluctuations as shown in Figure \ref{pulse_sequences}. The sequences on the left show occasional shifting in the pulse profile, corroborating the findings by \cite{CLH+20}. On the contrary, no such shifting is detected in the pulse sequence on the right. Careful examination also reveals visible movement of the subpulses, which results in a combination of regular and irregular subpulse tracks along the pulse sequences. In the following, we analyze and discern the emission between shifting and non-shifting.

	Emission during the events of shifting shows continuous changing in phase position to later longitudes, with an accompanying drop in the pulse intensity in the ``first half'' of the profile. To identify the events, we use data in Observation I and divide the integrated profile into two parts from its peak intensity, as indicated by the vertical dashed line in gray in Figure \ref{fig-Energy_criteria}. We refer to the longitudinal phase ranges in the parts of the profile to the left and to the right of the vertical line as the `leading' and `trailing' parts, respectively. The pulse energy is calculated using the following equation
	\begin{equation}
		E_p = \sum_{p=1}^n \eta[p, b:e].
	\end{equation}
	Here, $p$ represents a starting pulse number, and $n$ is the total number of pulses in the observation. The array $\eta$ contains the energy of a pulse within the on-pulse window whose boundary, in terms of bin number, begins at $b$ and ends at $e$. The identification of the on-pulse range is based on $10\%$ of the maximum intensity of the integrated pulse profile. Then, the total energy across the on-pulse window for a pulse at pulse number $p$ is stored in the array $E_p$. In addition, the values of $b$ and $e$ will change depending on the range of the profile under consideration. For example, when the leading part of the profile is concerned, the value of $b$ will have the bin number at the left boundary of the profile (at 10\% intensity) and $e$ has the bin number equivalent to the pulse phase at the peak of the integrated profile. We first determine a threshold value which is obtained from averaging the energy from all pulses in the leading part of the profile. Next, we execute a loop that ranges from 1 to 2785, with a step size of 30, to calculate the pulse energy for each successive block of 30 pulses. Figure \ref{fig-Energy_criteria} presents the pulse energy variation in Observation I. Pulses are characterized by their energy levels. A shifting pulse emission is recognized if the pulse energy in the leading part of the profile drops below the threshold, whereas a non-shifting pulse emission is signified by the pulse energy above the threshold. We have identified five shifting (shaded in gray) and six non-shifting emission modes in the observation.
	
\begin{figure*}
		\centering
		\includegraphics[width=1\columnwidth]{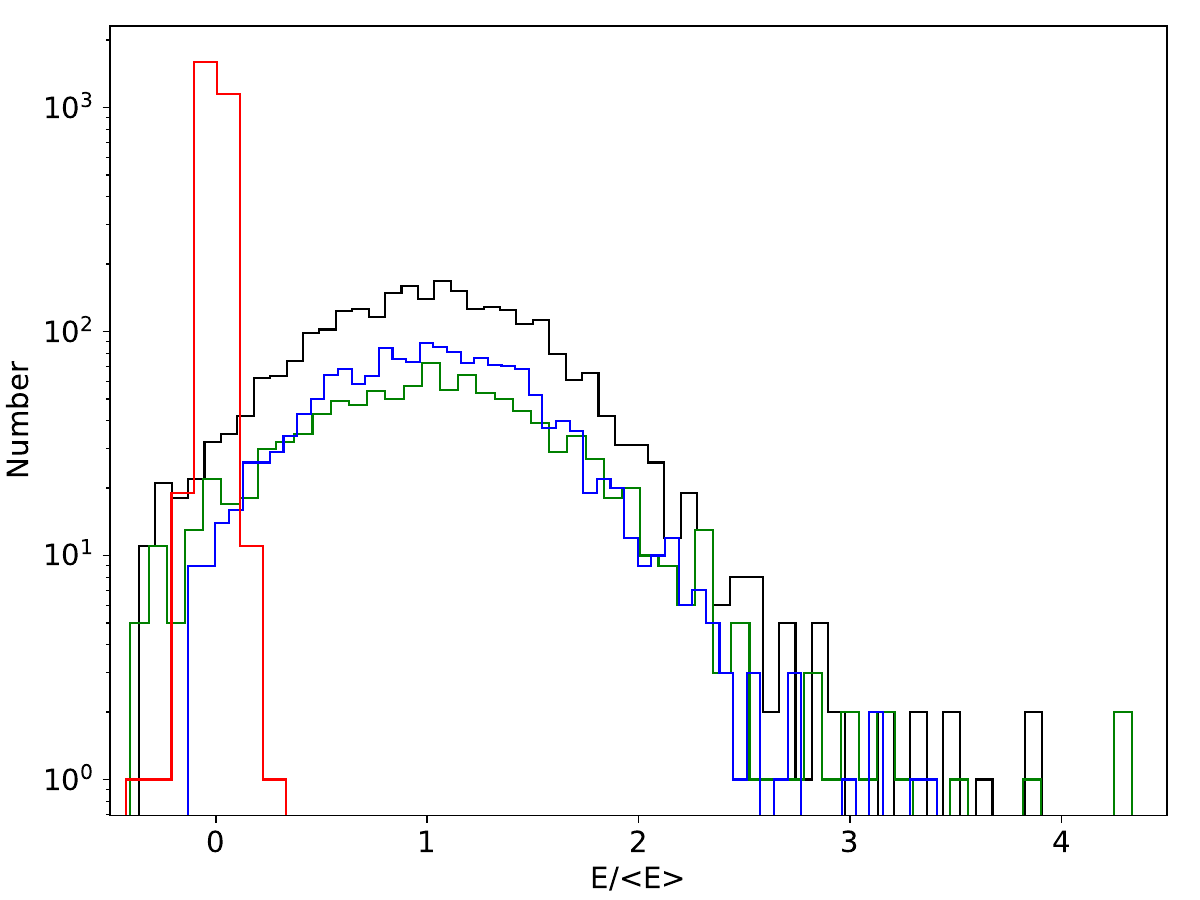} 
		\includegraphics[width=1\columnwidth]{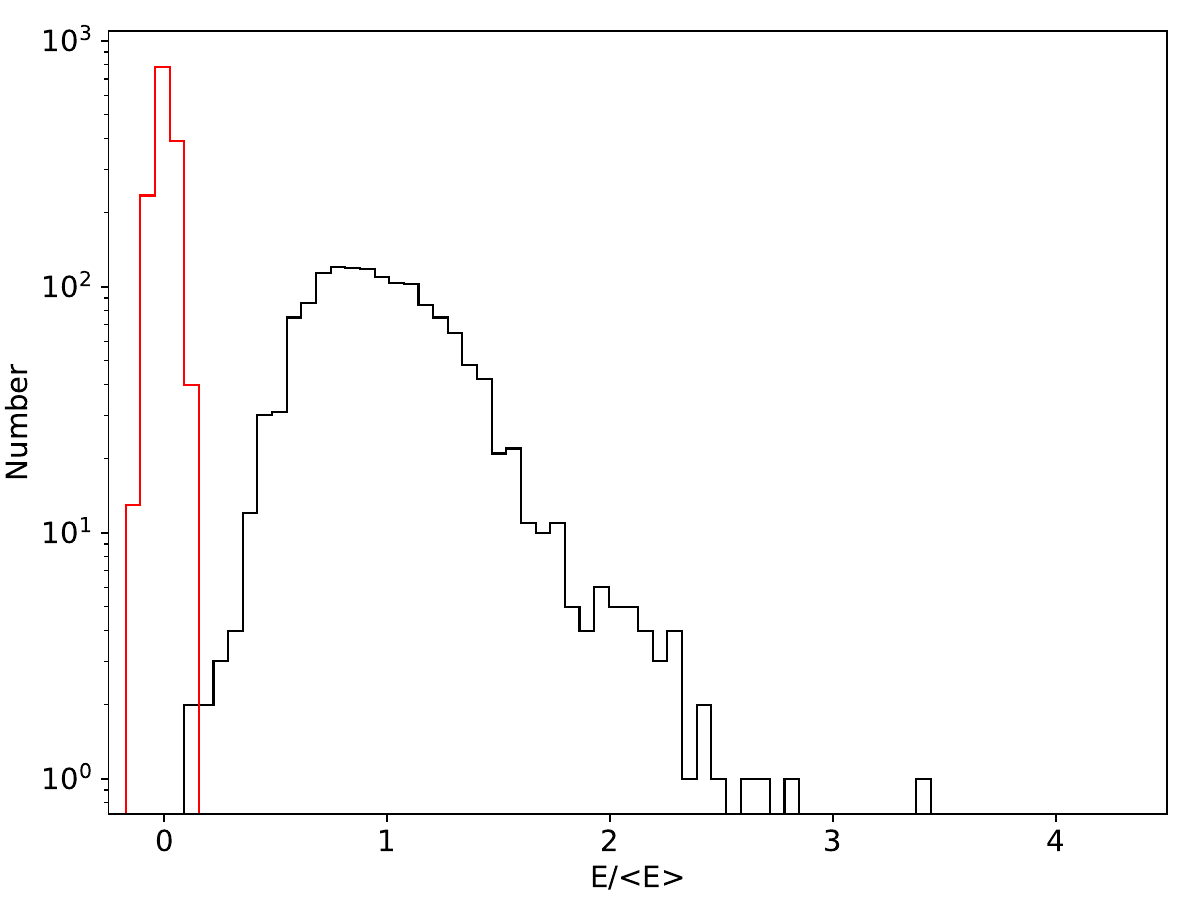} 
		\caption{The energy distributions for Observations I and II are shown in the left and right panels, respectively. The different colors represent the energy distributions for the on-pulse region (black), off-pulse region (red), shifting events (green), and non-shifting events (blue). Note that energy distributions for the on-pulse (black) and off-pulse (red) regions are shown for Observation II as shifting event is not present in the observation.}
		\label{fig-energy}
	\end{figure*}

	The pulse energy distributions for the on-pulse and off-pulse windows, as well as for the pulses during profile shifting, are shown in Figure \ref{fig-energy}. The on-pulse energy is computed for each pulse by adding up the intensities within the on-pulse region after subtracting the baseline noise. The off-pulse energy is calculated within the off-pulse region in a similar manner, with the same bin size as the on-pulse range. The histograms of the energy distributions are plotted by normalizing the $x$-axis with the mean on-pulse energy of the observation, and the $y$-axis is pulse number in logarithmic scale. Note that the on-pulse energy distribution for the pulses during profile shifting events is indicated in green. A small amount of on-pulse energy near zero is observed, which may be attributed to weak emission. Consequently, we believe that pulse nulling is not evident \citep{Ritchings76, Wang07} in the two observations.

	\section{Profile shifting}
	\label{sect:Analysis_sequences}
	
	In this section, we investigate the profile shifting as revealed in Observation I.
	
	\subsection{The shifting events}
	\label{sect:shifting}
	
	An instance of profile shifting is identified as an intensity drop in the leading part of the profile, and  the pulse emission shifts toward later longitudinal phases, where it remains for tens of pulse periods before returning to its original position. Our analysis reveals different durations for different shifting events, with the longest spanning 390 pulse periods whereas the shortest lasting only 90 pulse periods. In addition, the intervals between these events vary, with the longest being 330 pulse periods and the shortest just 150 pulse periods. To avoid misidentification with the intrinsic variation of the single pulses, only instances with more than two consecutive sub-integrations are included, which are marked by shaded areas in Figure \ref{fig-Energy_criteria}. On average, the duration for a shifting event is approximately 216 pulse periods, whereas non-shifting emission lasts for about 284 pulse periods. Shifting emission accounts for 38.78\% of the observation.

	\begin{figure}
		\centering  
		\includegraphics[width=\columnwidth]{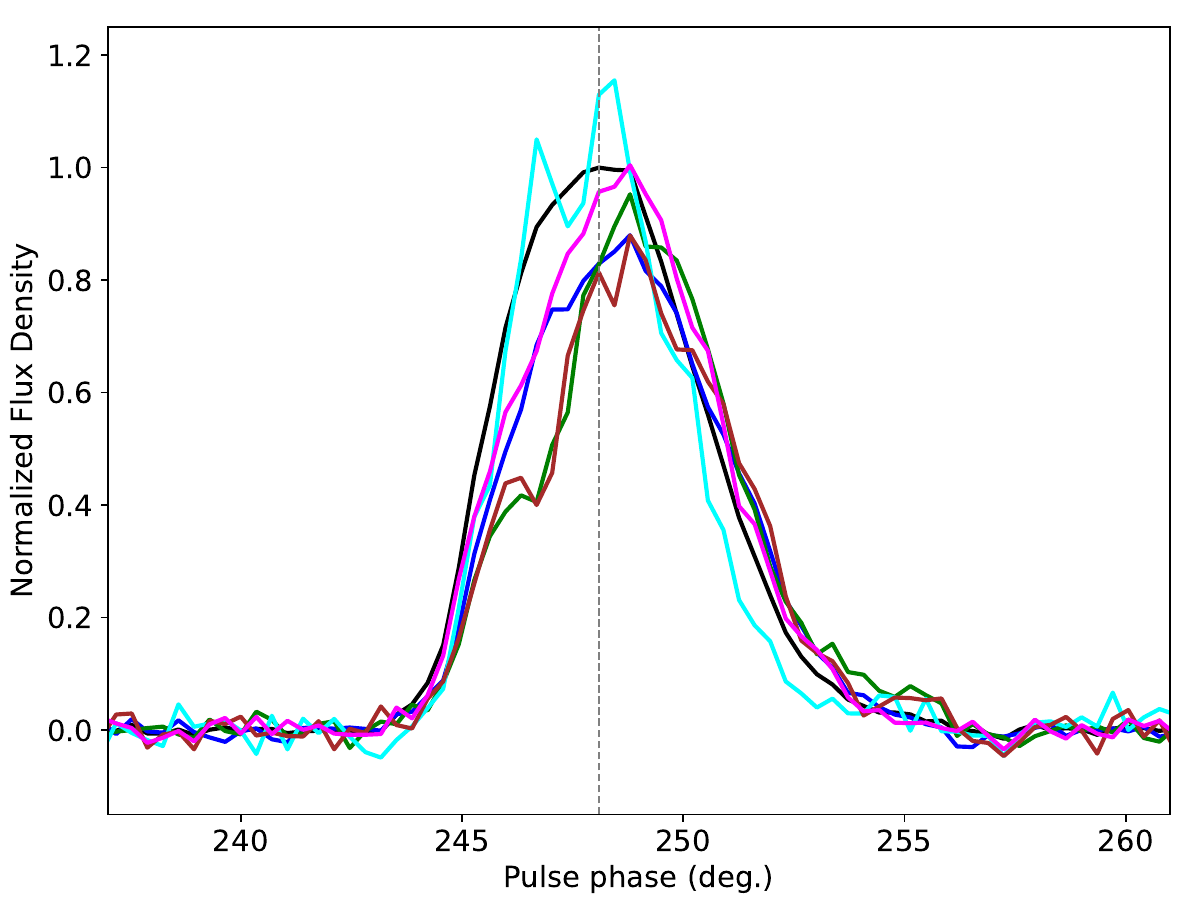}
		\caption{The integrated profiles for the shifting events are shown in blue (Event 1), green (Event 2), cyan (Event 3), brown (Event 4), and magenta (Event 5). The integrated profile obtained from Observation I is shown in black solid line, whose peak is indicated by the vertical dotted black lines, and all other profiles are normalized by its peak intensity.}
		\label{profile_shifting}
	\end{figure}
	
	\begin{table}
		\setlength{\tabcolsep}{4.5pt}
		\centering
		\caption{The details for each of the five shifting events in Observation I. Here, `Pulse \#' gives the pulse numbers where an event occurs, and its duration in pulse number is given in the column under `Span'. The separation between two consecutive events is indicated in the fourth column in units of pulse number. The values for $W_{10}$ and $W_{50}$ are measured in degrees.}
		\begin{tabular}{c|ccccc} 
			\hline
			Event & Pulse \# & Span & Separation & $W_{10}$ & $W_{50}$ \\ 
			\hline
			1 & 150--420 & 270 & 150 & 8.80 & 5.28\\
			2 & 660--810 & 150 & 240 & 9.15 & 3.87 \\
			3 & 960--1050& 90 & 150 & 7.04 & 4.22\\
			4 & 1380--1560 & 180 & 330 & 8.45 & 4.93\\
			5 & 1830--2220 & 390 & 270 & 8.80 & 4.93\\
			\hline
		\end{tabular}
		\label{table:shifting}
	\end{table} 
	
	\begin{figure}
		\centering  
		\includegraphics[width=1\columnwidth]{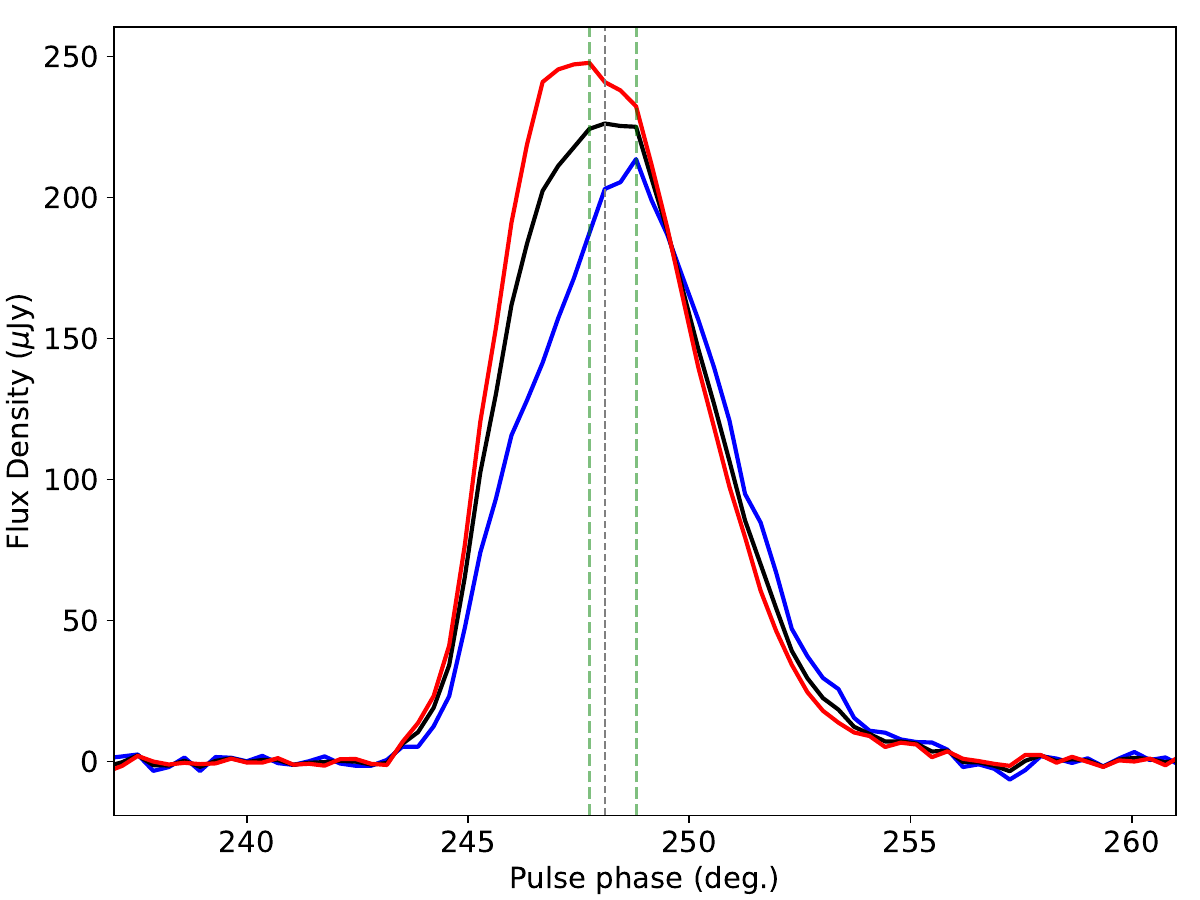}
		\caption{Plot showing the integrated profiles obtained from pulses during shifting (blue) and non-shifting (red), and the profile obtained from all the pulses from Observation I is shown in black. Green dashed lines mark peak intensities for non-shifting (left) and shifting (right), while the dotted black line represents the peak intensity in the Observation I profile.}
		\label{shifting_nonshifting}
	\end{figure}
	
	The emission characteristics observed across various profile shifting events exhibit marked differences. In addition to different duration, the integrated profiles for the five events relative to that from the entire observation are also different as shown in Figure \ref{profile_shifting}. There is an apparent shift of approximately $0.7^\circ$ measured from the profile in black to the peaks of other profiles. We found that the profile from Event 3 displays the highest peak intensity, while the profile from Event 5 possesses a peak intensity nearly equal to the profile in black. However, the peak intensities for the profiles in the remaining events are all lower than the peak intensity of the profile in black. Variation in profile width is also seen among the shifting events as shown in Table \ref{table:shifting}. Measured at 10 percent intensity ($W_{10}$), Event 2 has the broadest pulse width, with about 4.14\% more than the profile in black, whereas Event 3 has the narrowest profile width. The profile width from Event 4 mostly matches that from the entire observation, and profiles from both Events 4 and 5 align with the entire observed profile. However, larger changes are seen at 50 percent intensity ($W_{50}$). Profile from Event 1 demonstrates the widest width, having increased by about 7.1\% compared to the black profile. Event 2 is the narrowest, with a decrease of 21.50\% compared to the entire observed profile.

	Figure \ref{shifting_nonshifting} shows the integrated profiles for the non-shifting emission pulses (red), the shifting events (blue), and for the entire observation (black). The shift between the peak intensities of the non-shifted (left) and shifted (right) emission profiles (between the two green dashed lines) is approximately $1^\circ$ in longitude. In addition, the peak of the non-shifting profile, which is the strongest among the three profiles, is located $0.3^\circ$ earlier in longitudes relative to the peak of the integrated profile from the entire observation period. However, the peak intensity of the shifted profile is lower than that of the non-shifted profile by about 10\%. In addition, the shifted profile clearly reveals more emission components than the non-shifted counterpart. The variations in profile width the solid black, blue, and red profiles are, respectively, $8.45^\circ$, $8.8^\circ$, and $8.09^\circ$ at $W_{10}$ and $4.93^\circ$, $4.93^\circ$, and $4.57^\circ$ at $W_{50}$. While the changes are approximately consistent across the three profiles at the two intensity levels, the changes in the blue profile relative to the black profile become more pronounced as the intensity level increases to $W_{75}$ and higher.
	
	Profile shifting has been detected in a few radio pulsars. \cite{RRW06} discovered the first profile shifting event, later referred to as `swooshing', in PSRs J1901$+$0716 (B1859$+$07) and J0922$+$0638 (B0919$+$06) using the 305-m Arecibo telescope. The profiles appear to displace about $5^\circ$ from their original locations to earlier longitudinal phases together with an increase in the profile width. However, the profile shifting in PSR J0344$-$0901 is different in that the profile tends to shift toward later longitudinal phases. \citet{CLH+20} reported that the event was present in two out of three observing sessions using the FAST telescope. We detected the shifting events only in one of two observations, which made up about two thirds of the total observing time. In addition, the fact that the shifting event did not occur at all in Observation II, whose duration is longer than the average separation between two consecutive shifting events in Observation I, would imply that the event likely takes place in groups.
	
	\subsection{Changes in the polarization and the position angle}
	
\begin{figure}
		\centering
		\includegraphics[width=1\columnwidth]{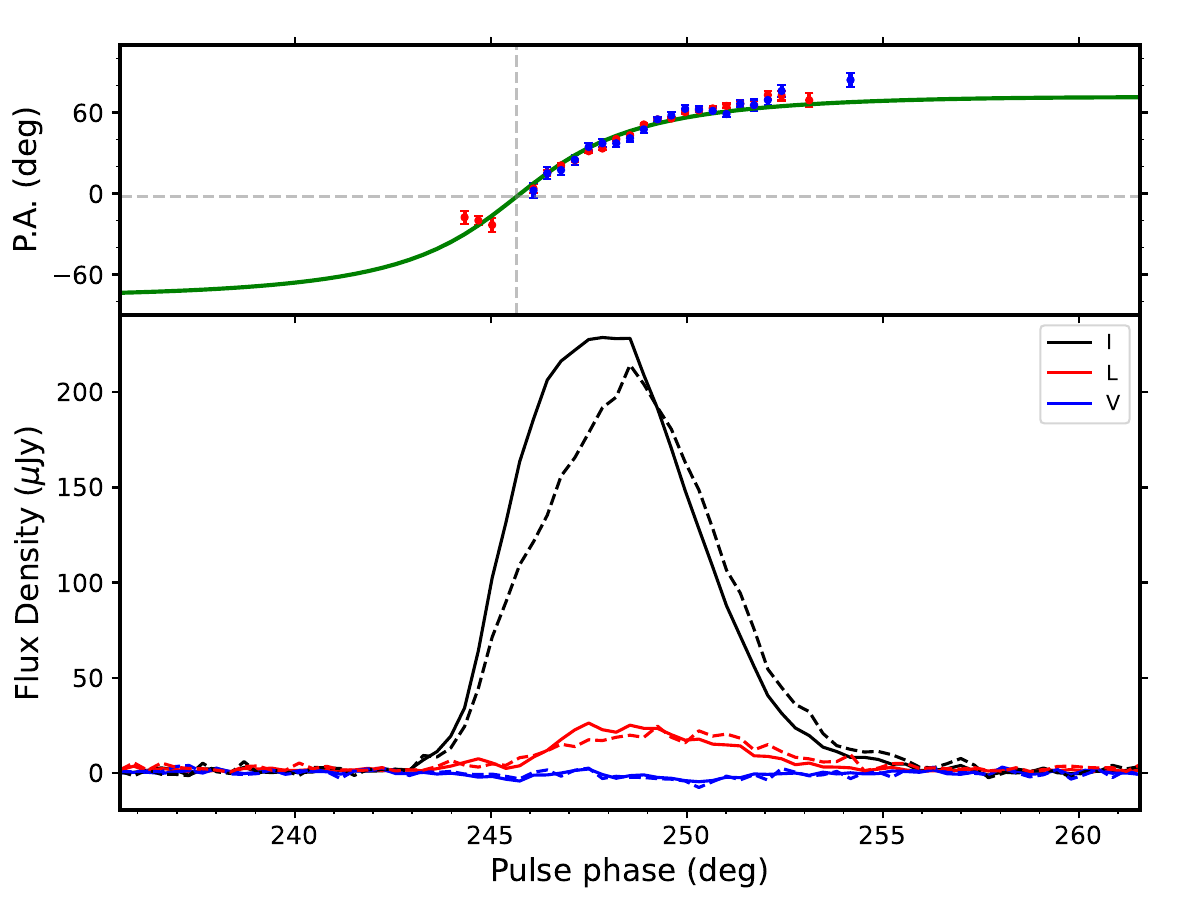}
		\caption{Changes in the polarization and the position angle from Observation I. The upper panel shows the changes in the polarization position angle (PPA) in red and blue across, respectively, profiles from all the pulses (black solid in the lower panel) and from the shifting events (black dashed in the lower panel). The curve in green represents the best fit to the PA swing, and the crossing between the vertical and horizontal gray dashed lines illustrates the point of the steepest gradient (SG) of the RVM curve. In the lower panel, the linear (L) and circular (V) polarizations across the two profiles are shown.}
		\label{fig-profiles_polarizations}
	\end{figure}

	Variations in the polarization and the position angle across the shifted profile as compared with that from the whole observation are shown in Figure \ref{fig-profiles_polarizations}. In the lower panel of the figure, the Stokes parameters $I$ (black), the linear polarization $L=\sqrt{Q^{2}+U^{2}}$ (red), and the circular polarization $V$ (blue) are shown for the profiles obtained from all the pulses and from the shifting events. The polarization position angle, $\psi$, is calculated using $\psi =\tan^{-1}(U/Q)/2$, where $U$ and $Q$ are Stokes parameters for the two directions of linear polarization. The pulsar possesses only modest linear polarization. During profile shifting, the circular polarization remains mostly unchanged whereas the peak of the linear polarization shifts slightly towards later longitudinal phases. In addition, the position angle exhibits a noticeable change around the trailing half of the shifted profile as compared with that across the same section of the profile from all pulses. The overall range of pulse phase for changes in the PA is also shortened from the leading edge of the shifted profile resulting in a lesser symmetric curve.
	
		\begin{figure*}
		\centering  
		\includegraphics[width=2\columnwidth]{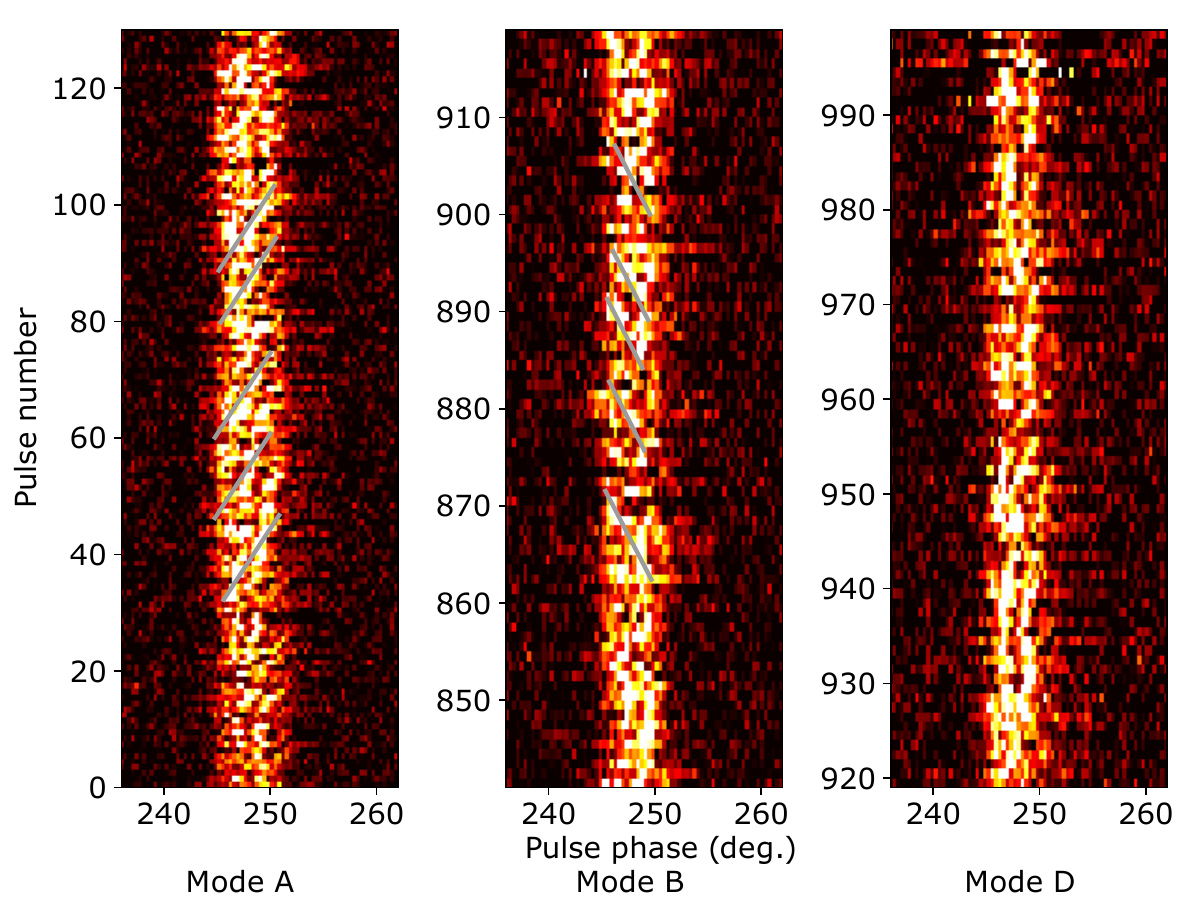}
		\caption{Samples of pulse sequence containing the different emission features observed in Observation I. The modes A, B and D are shown, respectively (see main text for description of the different emission modes). The gray lines illustrate the apparent movement of the subpulses.}
		\label{fig-drift_modesI}
	\end{figure*}
	\begin{figure*}
		\centering  
		\includegraphics[width=2\columnwidth]{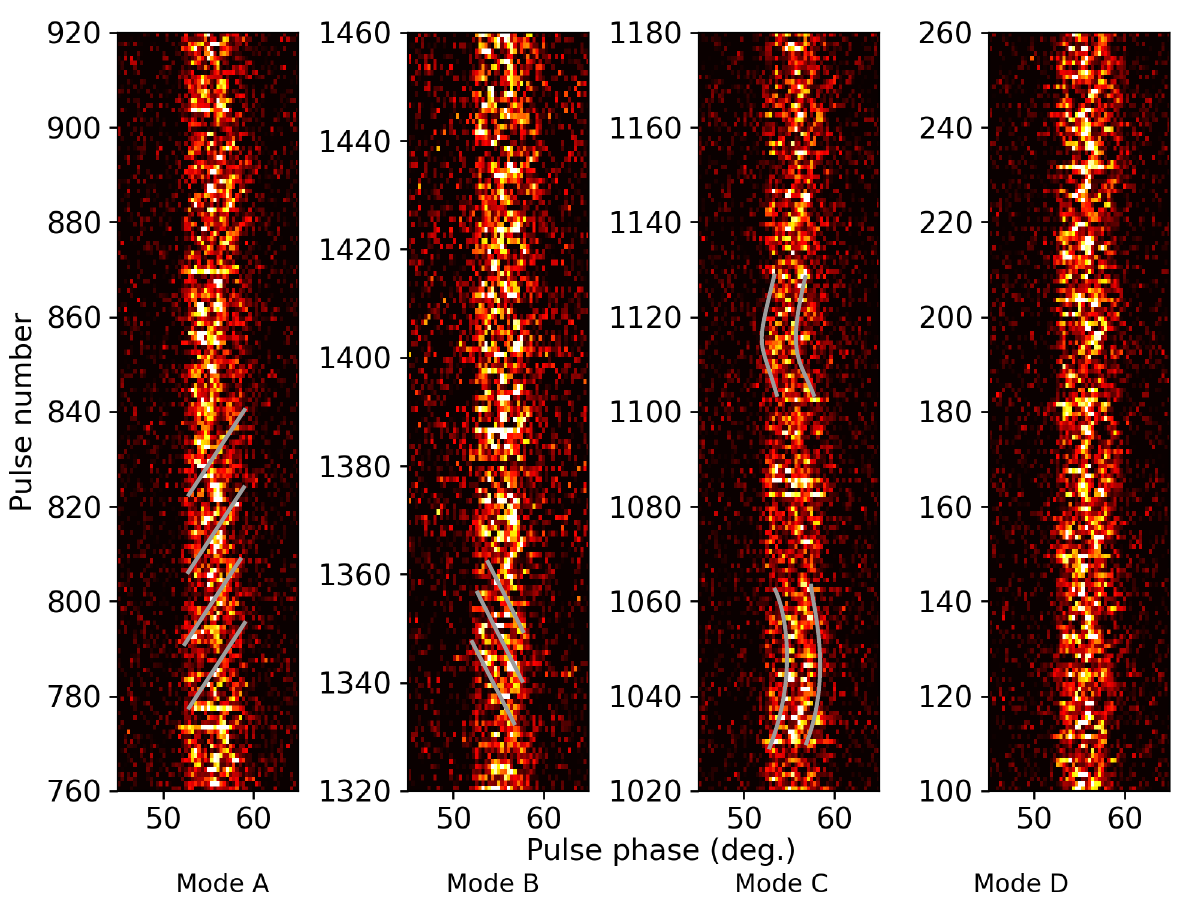}
		\caption{The different emission modes in Observation II. From left to right are the modes A, B, C and D, respectively. The gray lines illustrate the apparent movement of the subpulses.} 
		\label{fig-drift_modesII}
	\end{figure*}
	
	The swing of the position angle across the integrated profile of the entire observation is close to an $S$-shape, as shown by the red dots in the upper panel of Figure \ref{fig-profiles_polarizations}. Therefore, we assume the rotating vector model \citep[RVM;][]{RadhakrishnanCooke1969} for fitting the PA swing. The polarization position angle ($\psi$) of the linearly polarized emission can be expressed as a function of pulse phase $\phi$ \citep{Lyne88},
	\begin{equation}
		\tan(\psi-\psi_{0}) = \frac{\sin\alpha\sin(\phi-\phi_{0})}{\sin\zeta\cos\alpha - \cos\zeta\sin\alpha\cos(\phi-\phi_{0})}.
	\end{equation} 
	Here, $\alpha$ represents the angle between the magnetic and the rotation axes, and $\zeta$ signifies the angle between the line of sight and the rotation axis. We obtain the best-fit parameter of $75.12^\circ \pm 3.80^\circ$ for $\alpha$ and $71.95^\circ\pm 3.73^\circ$ for $\zeta$, with the reduced $\chi^2$ being 4.02. The vertical and the horizontal gray dotted lines at $\phi_{0}$ = $246.50^\circ  \pm 0.14^\circ$ and $\psi_{0}$ = $10.70^\circ \pm 2.54^\circ$ in the upper panel represent the maximum steepest gradient (SG) of the RVM curve given by
	\begin{equation}
		\Big(\frac{d\psi}{d\phi}\Big)_{\rm max} = \frac{\sin\alpha}{\sin\beta},
	\end{equation} 
	where $\beta$ = $\zeta -\alpha = -3.17^\circ \pm 5.32^\circ$ is the impact parameter of the pulsar. From the fitting using the RVM model, the negative $\beta$ corresponds to the positive gradient in the PA swing. The SG point is estimated at approximately $-17.48^\circ \pm 29.34^\circ$, which is located near the leading edge of the profile. The location of the SG point at one edge of a profile is the characteristic of `partial cone' pulsars as identified by \citet{Lyne88}. These pulsars possess profiles that show one steeply rising edge and another slowly falling edge, and also feature a PA traverse with SG point positioning toward one edge of the profile. Typically, the SG point on the `S-shaped' curve, which indicates the emission geometry of a pulsar, coincides with the profile center in long-period pulsars \citep{Lyne88}. However, in PSR J0344$-$0901, the non-central SG point suggests that it may be a `partial cone' pulsar. The influence of aberration and retardation effects on the emission geometry have been identified in slower rotating pulsars \citep{Blaskiewicz1991, Mitra11}. Given the large inclination angle of our pulsar, the magnetic field in the emission region is likely distorted from a purely dipolar structure and the shape of the polar cap becomes asymmetric \citep{CRZ00}. Assuming emission originating from the last open dipolar field-lines, the estimated emission height is 288.12 kilometers measured from the center of the star, or about 0.5\% of the radius of the light cylinder. This indicates that for long-period pulsars with large inclination angle, such as PSR J0344$-$0901, the impact of aberration and retardation effects can still be significant.

	\section{The emission variation}
	\label{sect:drifting}
	
	In this section, we explore the emission variation detected in Observations I and II.

	\subsection{The different emission features}
	
	An inspection of the pulse sequence in Figures \ref{fig-drift_modesI} and \ref{fig-drift_modesII} reveals significant changes in the longitudinal phase where the subpulse emission is detected. This gives rise to variation in the tracks traced by the subpulses in the pulse sequence. We categorize the emission features into four distinct modes, each is characterized by unique pattern of the subpulse tracks.
	
	\begin{figure*}
		\centering
		
		\includegraphics[width=1\columnwidth]{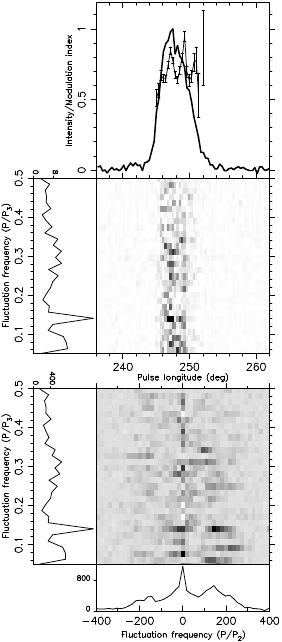} 
		\includegraphics[width=1\columnwidth]{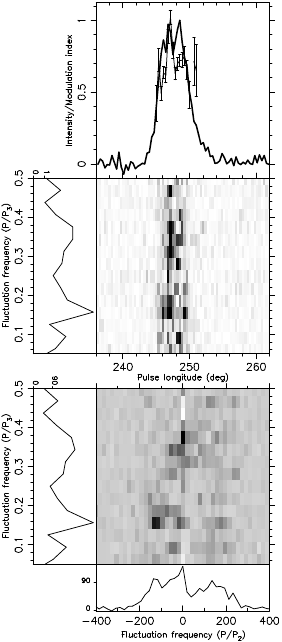} 
		
		\caption{Example pulse sequences showing the fluctuation analysis for Modes A (left column) and B (right column) from Observation I. Top: Pulse profiles with modulation index. Middle: LRFS with side panel showing horizontally integrated power, highlighting differences between Modes A and B. Bottom: 2DFS for each mode, with side panels indicating horizontal and vertical integrated power. The pulse sequence for Mode A is taken from pulse numbers between 20 and 130 in total of 110 pulses. A 64-point Fourier transform is employed, and then averaged over the blocks of the entire pulse sequence. Similar for Mode B, which encompasses 40 pulses from pulse numbers 840 to 880, a 32-point Fourier transform is employed, and also averaged over the blocks of the whole pulse sequence.}
		
		\label{Modes_I}
	\end{figure*}
	
	\begin{figure*}
		\centering
		
		\includegraphics[width=1\columnwidth]{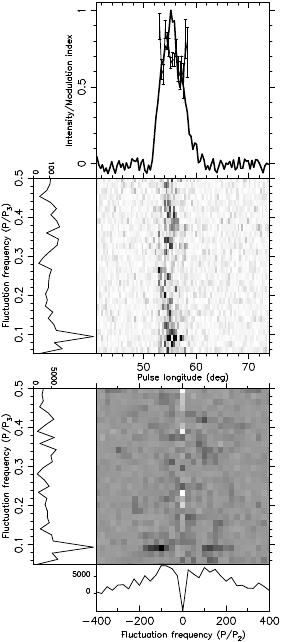} 
		\includegraphics[width=1\columnwidth]{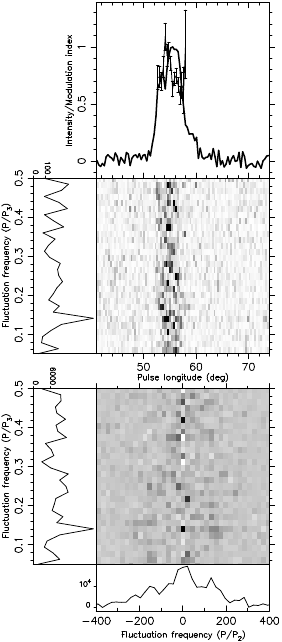} 

		\caption{Similar to Figure \ref{Modes_I}, the plots present the LRFS and 2DFS from Observation II. Distinct periodic modulations in the pulsar's emission are evident in both Modes A (left column) and B (right column). An example of Mode A spanning 110 pulses is taken from pulse number 790 to 900, while Mode B captures 59 pulses from 1381 to 1440. In the top panel in each plot, the integrated power over frequency is depicted. A 64-point and 32-point Fourier transform were employed and averaged over the blocks of the entire pulse sequence for Modes A and B, respectively.}
		\label{Modes_II}
	\end{figure*}

	\noindent{\it Mode A}:
	The tracks appear tilted consistently towards later longitudinal phase. An example is shown in the leftmost pulse sequence in Figures \ref{fig-drift_modesI} and \ref{fig-drift_modesII}, in which the subpulse tracks are indicated by the gray lines showing the direction of the subpulse movement. In addition, the subpulse movement can be seen across almost the whole profile, and the slope of the tracks may or may not be constant. Furthermore, variation in the track separation is also observed, as shown between pulse numbers 20--130 in Observation I and 760--920 in Observation II, with the tracks in the former being closer to each other. This demonstrates that changes in subpulse movement occur even within the same mode. This mode is the second most abundant in our observation, which covers around 42\% and 28\% of the single pulses in Observations I and II, respectively.

	\noindent{\it Mode B}: 
	Similar to Mode A, the subpulse tracks in this mode are also seen across the whole profile except that they appear tilted toward earlier longitudinal phases. This indicates that the subpulse movement is also toward the same direction. Examples are illustrated in the middle plot in Figure \ref{fig-drift_modesI} and the second plot (from the left) in Figure \ref{fig-drift_modesII}. This mode covers nearly 14\% and 11\% of the single pulses in Observations I and II, respectively, making it the third common emission mode in the observations.
	
	\noindent{\it Mode C}:
	The subpulse pattern in Mode C differs distinctly from that in both Mode A and Mode B. The subpulse tracks in this mode exhibit a continuous change in direction, resulting in a curved pattern. This is apparent in the third pulse sequence (from the left) in Figure \ref{fig-drift_modesII}. The tracks can be short, as shown between pulse numbers 1030 and 1060, or long, as observed between pulse numbers 1090 and 1150. This mode occurs less frequently, representing about 6\% of the single pulses. Subpulse movement in this mode occurs across the entire profile, similar to Modes A and B, and changes in the direction generally take place concurrently across adjacent tracks. Mode C was detected only in Observation II, indicating that it may not be a prevalent mode in this pulsar.
	
	\noindent{\it Mode D}:
	This mode is marked by irregular track patterns that do not correspond to any of the three categories above. The subpulses appear to move randomly in both directions or follow a curved path without recurring patterns. Examples are provided in the rightmost pulse sequence in Figures \ref{fig-drift_modesI} and \ref{fig-drift_modesII}. Despite these irregularities, Mode D represents the most common subpulse movement in the two observations covering 44\% and 55\% in Observations I and II, respectively.

	\subsection{Periodicity of the subpulse track patterns}
	
	Since the subpulse tracks are repeating in modes A and B, it is possible to use the phase-averaged power spectrum (PAPS) method \citep{SMK05, Wen16, BM18} to examine the subpulse movement. The method measures the vertical separation between consecutive tracks, which is also commonly designated as $P_3$ in subpulse drifting. Subpulse drifting demonstrates as systematic and (usually) long-lasting marching of subpulses across the pulse window \citep{Pulsars1977, Weltevrede06}. In the following, we also refer to the separation between subpulse tracks as $P_3$. Note that the continuously changing subpulse tracks in Mode C pose a challenge for identifying the associated $P_3$ values, and hence we exclude them from the following calculation. For a selected pulse sequence, Fourier transform is performed to determine the absolute values of the pulse flux density at each longitudinal phase across the pulse window. We then compute the PAPS value by averaging the resulting transforms over the pulse phase, which yields a frequency resolution that ranges from 0 to 0.5 cycles per pulse period, $c/P$,  which is equivalent to $P/P_{3}$. Our analysis of fluctuation spectra indicates that any low-frequency structure observed in the PAPS, with a value less than 0.05, can be considered negligible. We then obtain the $P_{3}$ value by taking the reciprocal of the significant peak value from the frequency resolution. Finally, we utilized the Longitude Resolved Fluctuation Spectrum  \citep[LRFS;][]{Backer1973}, akin to PAPS, to characterize and estimate $P_{3}$.
	
	Our observations clearly depict periodic modulations in both Mode A and Mode B, as showcased in the two example pulse sequences in Figures \ref{Modes_I} (Observation I) and \ref{Modes_II} (Observation II). These modes are prominently displayed in the middle panel of each plot, highlighting periodic characteristics with a distinct surplus of power within the respective pulse phase range. The vertical axis on the left panel of each plot indicates $P/P_{3}$. For both modes, we observed that $P_3$ is not consistent and exhibits varying values across different pulse sequences. The upper panel of each plot presents the pulse profile, accompanied by a comprehensive depiction of the modulation index for Modes A and B in both Observations I and II. This modulation index measures the pulse intensity fluctuation and can be mathematically expressed as $m_i$ = $\sigma_{i}/\mu_{i}$ \citep{Weltevrede06}. Here, $\sigma_{i}$ represents the longitude-resolved standard deviation, while $\mu_{i}$ is the mean intensity for the designated longitude bin $i$. This metric provides insights into pulse intensity variability at particular pulse longitudes based on the LRFS methodology \citep{Edwards03, Weltevrede12}. Error bars for the modulation index are determined through bootstrapping the data. Each bootstrap iteration introduces random noise corresponding to the rms of the off-pulse region.

\begin{figure}
		\centering  
		\includegraphics[width=1\columnwidth]{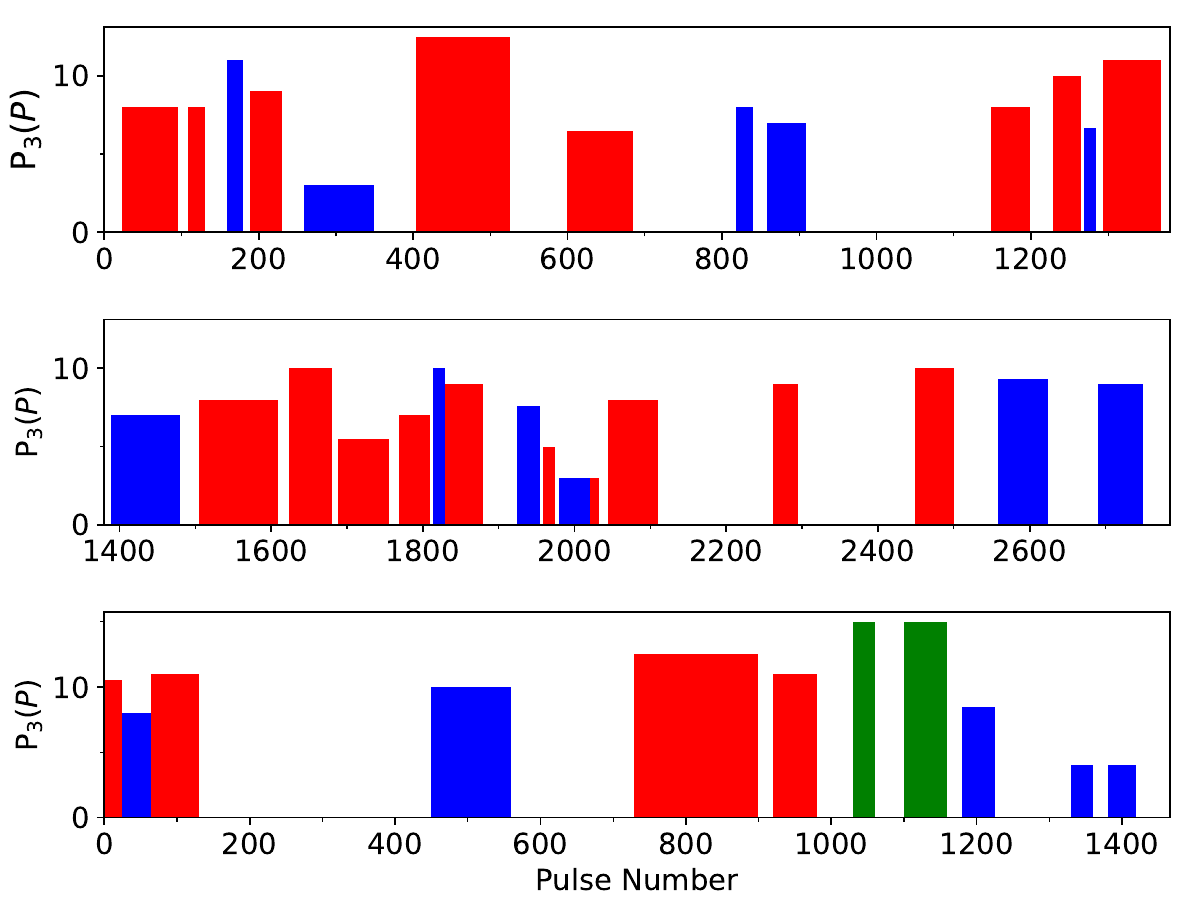}
		\caption{Distribution of $P_{3}$ values for Observation I (upper and middle plots) and Observation II (lower plot). Modes A and B are shown by red and blue rectangles, respectively. Mode C is shown in green using a guessed $P_{3}$ value. Empty areas indicate Mode D, without a well-defined $P_3$ value.}

\label{P3_perodicity}
\end{figure}

	For Observation I, as shown in Figure \ref{Modes_I}, the modulation displays a peak frequency at roughly 0.14 $P/P_{3}$ for Mode A, while it is approximately 0.16 $P/P_{3}$ for Mode B. The pattern periodicity for the two modes corresponds to $P_{3}$ $\sim$ 7.14 $P$ for Mode A, and $P_{3}$ $\sim$ 6.25 $P$ for Mode B. After conducting a comparative analysis, as shown in Figure \ref{P3_perodicity}, the mean $P_3$ values for Modes A and B in Observation I are approximately 8.73 $\pm$ 0.5$P$ and 7.88 $\pm$ 0.46$P$, respectively. As for Observation II, as depicted in Figures \ref{Modes_II}, the modulation for Mode A indicates a peak frequency of approximately 0.1 $P/P_{3}$ or $P_{3}$ $\sim$ 10 $P$. For Mode B, the frequency is around 0.14 $P/P_{3}$ or $P_{3}$ $\sim$ 7.14 $P$. Likewise, in Observation II, the average $P_3$ values for Modes A and B are around 11.76 $\pm$ 0.11$P$ and 7.88 $\pm$ 0.3$P$.

\begin{figure}
\centering  
\includegraphics[width=1\columnwidth]{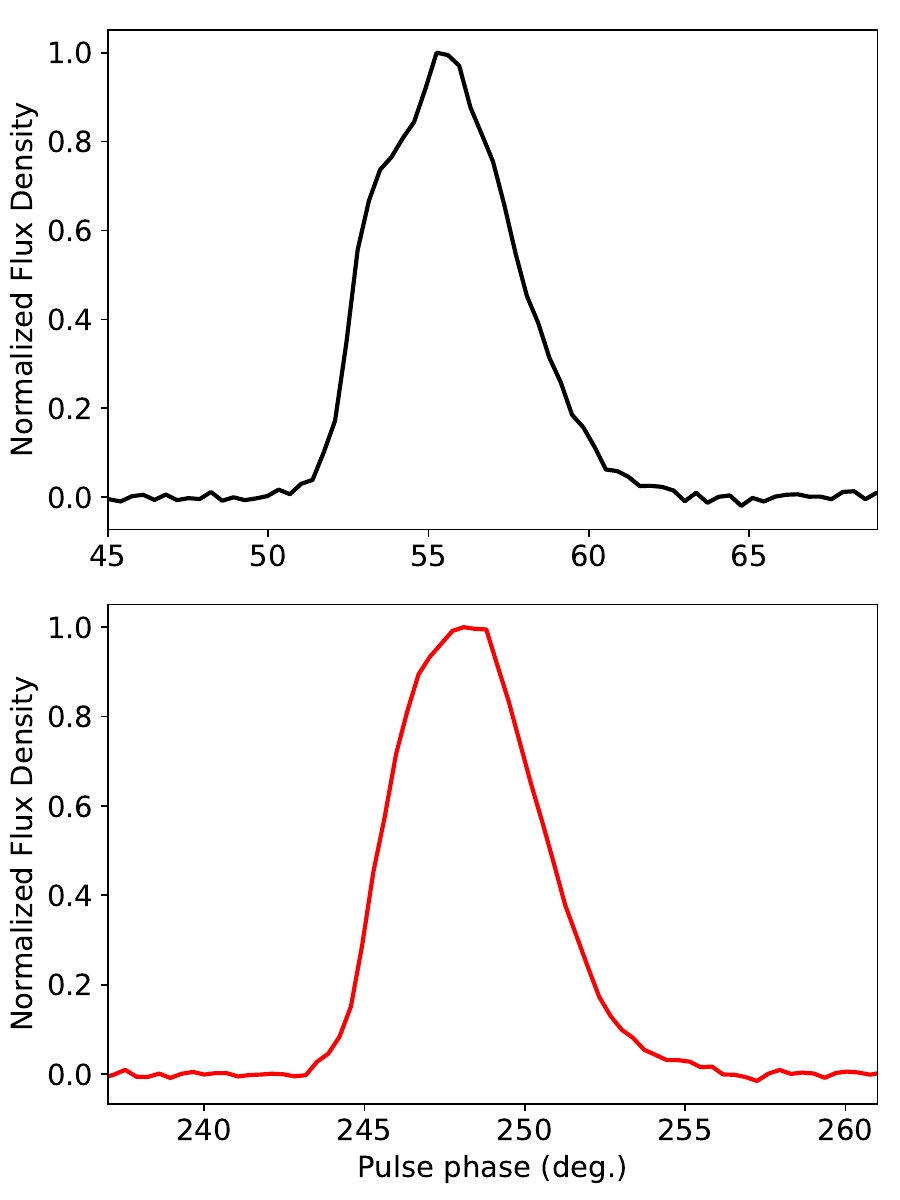}
\caption{The normalized integrated profiles from Observation II (black) and Observation I (red).}

\label{profile_differ}
\end{figure}

\begin{figure}
\centering  
\includegraphics[width=1\columnwidth]{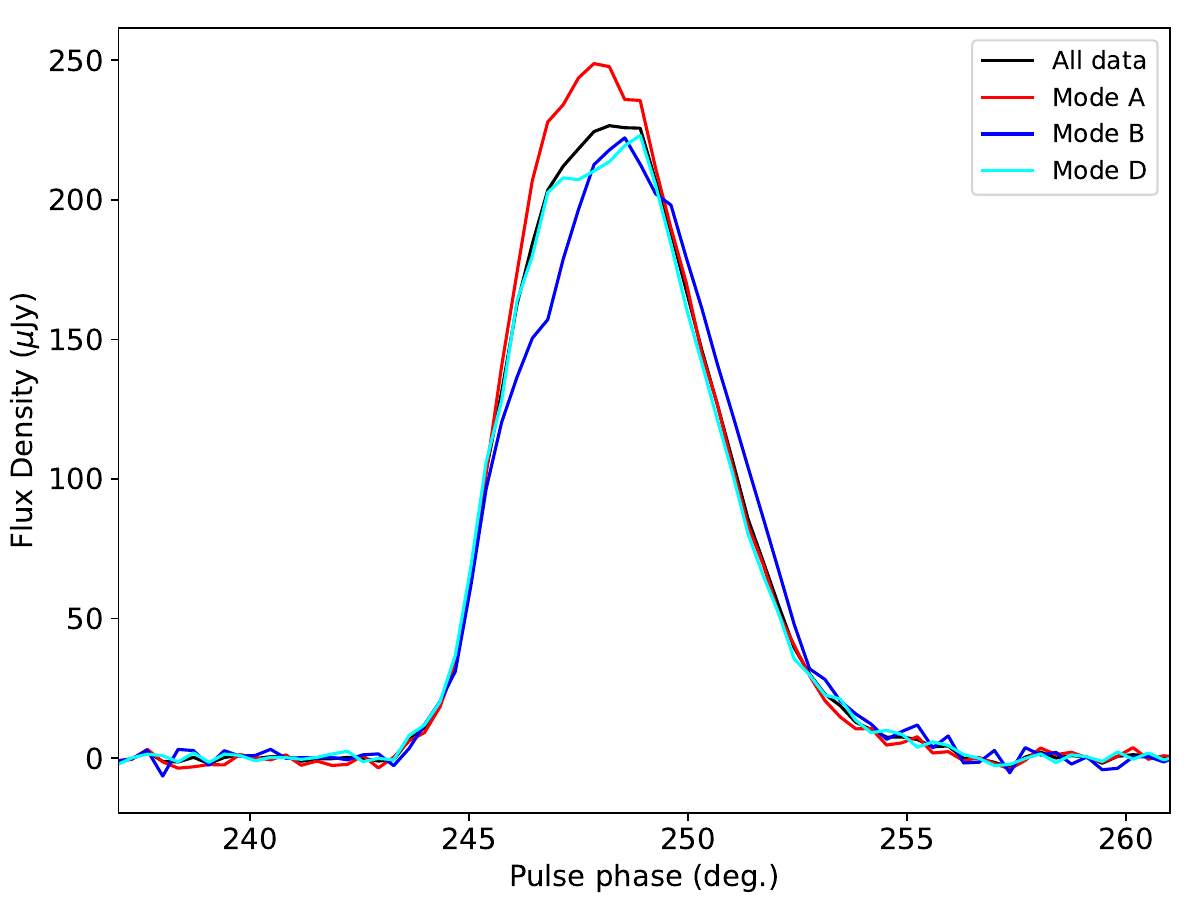}
\includegraphics[width=1\columnwidth]{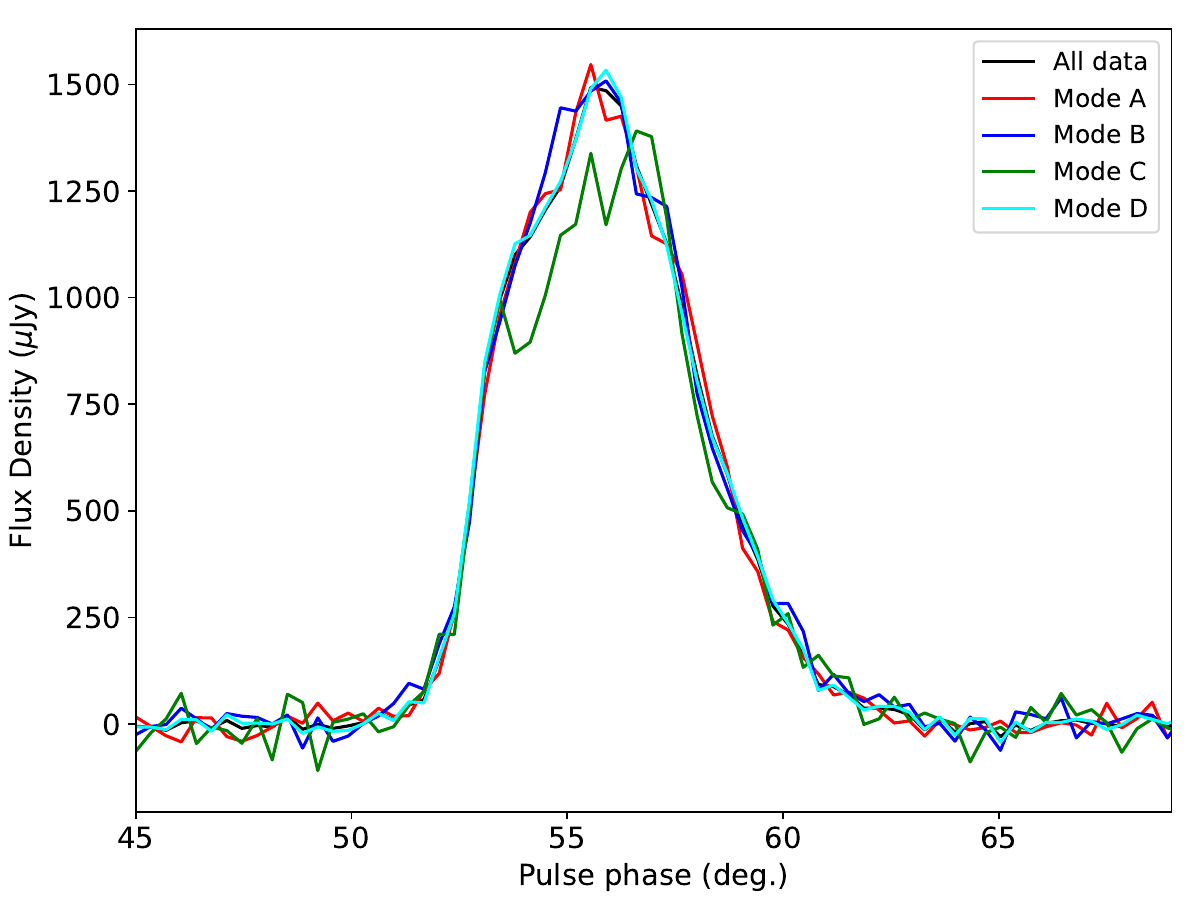}
\caption{The integrated profiles for the different modes obtained from Observation I (upper panel) and Observation II (lower panel).}

\label{profile_drifting}
\end{figure}

\subsection{Emission mode transition}

The different emission modes each manifested with particular $P_3$ value. The distribution of $P_{3}$ values for the two observations are shown in Figure \ref{P3_perodicity}. It demonstrates that changes from one mode to another are common throughout the two observations. However, such transitions appear random without showing discernible orders. Furthermore, the transitions do not accompany with any nulls. We found that Mode A possesses the longest duration in both observations. The longest duration reaches 121 pulse periods, and the shortest duration is 18 pulse periods. Nevertheless, the duration for this mode fluctuates as indicated by $P_3$. Mode B occurs less frequently, with the duration ranging from 23 to 90 pulse periods giving an average duration of 42 pulse periods. As for Mode C, its track pattern is less periodic than Mode B. 

\subsection{Profile variation}

The integrated profiles obtained from all pulses in Observations I and II is shown in Figure \ref{profile_differ} for comparison. The numbers of pulses that formed the two profiles are 2785 (Observation I) and 1466 (Observation II). It is generally suggested that an integrated profile with a stable shape could be obtained with about 500 pulses \citep{Pulsars1977}. However, the number of pulses required for such stabilization has also been shown to relate to the observed emission features \citep{HMT75}. In the case of PSR J0344$-$0901, the emission variation is significant, so that a longer stabilization time is required. This is evident from the different profile shapes shown in Figure \ref{profile_differ}, even though the number of pulses used to form each profile is a few times more than that suggested.	

It is expected that the different modes of subpulse movement will have impacts on the profile morphology.  When delving into the profiles at $W_{10}$, a uniformity in the profile width at $W_{10}$ is evident across all modes, as shown in Figure \ref{profile_drifting}. This is consistent with the argument that certain width parameters may remain relatively consistent across different modes due to the overarching pulsar emission geometry \citep{BM18}. However, this consistency disrupts at $W_{50}$, where the profile of Mode B is noticeably narrower. In addition, Mode A consistently displays the highest peak intensity in both observations. This agrees with previous studies that highlights the prominence of certain modes over others due to intrinsic pulsar properties \citep{SMK05}. In contrast, Modes B and D show similarities in their peak intensities. However, Mode C possesses an intensity that is markedly lower than the other modes.

\section{Discussion}
\label{sect:discussion}

In this section, we discuss similar events detected in other pulsars, and comment on several properties of the two phenomena in relation to the traditional models for drifting subpulses.

\subsection{Emission variations and the subpulse movement}

In the traditional models, subpulse drifting is described as originating from subbeams located on a carousel that is rotating at a rate determined by the $\vec{E} \times \vec{B}$ drift \citep{RudermanSutherland1975}. The subbeams rotate through the fixed line of sight giving rise to a systematic pattern of subpulse movement across the profile window with the drift characteristics being different for different pulsars. The model is effective for interpreting the emission geometry and subpulse drift behavior in many pulsars \citep{Bhattacharyya2007, Rankin2017, BM18}. In this model, the subpulse drift-rate and drift direction of a pulsar do not change meaning that the carousel rotates consistently. However, the model is under challenge as increasing number of pulsars have been found to exhibit unusual drifting subpulses demonstrating as changes between different drift-rates with such changes being pulsar-specific. It shows that the subpulse drift-rate of these pulsars is time dependent. To address these issues, alternative models have been proposed. This includes the suggestion of variation in the number of subbeams on the rotating carousel in different drift regions leading to different emission modes \citep{Gupta2004, Bhattacharyya2009, McSweeney2019, McSweeney2023}. Other proposals involve the partially screened gap model \citep{Gil2003, Szary2015} and the ion-proton radio pulsar polar cap model \citep{Jones2020}. There are also proposals for changes in the magnetospheric geometry as the cause for the observed changes in the emission properties \citep{Timokhin2010}, and the introduction of multiple emission states in the pulsar magnetosphere and switching between different emission states corresponds to changes in the electric drift resulting in the different subpulse drift patterns \citep{Yuen19}. Nevertheless, the rotating carousel is assumed in most of these models. Since drifting subpulses are closely related to the emission properties in the emission region, the different subpulse movements in PSR J0344$-$0901, and its variations, as revealed by the different $P_3$ values, indicate that the rotation of the carousel is not fixed but varies with time. It may be that the magnetosphere of PSR J0344$-$0901 also contains different emission states each with a unique rotation rate for the carousel. Then, a switch in the rotation rate will result in a change in the subpulse drift pattern leading to the observed variation in the subpulse movement and the profile shifting.

Subpulse movement is visible throughout our observations, even before and after a profile shifting, as shown in the left panel of Figure \ref{pulse_sequences}. In the traditional models, subpulses drift when a relative motion exits between the plasma flow and the corotation in the magnetosphere. The variation of the subpulse movement in PSR J0344$-$0901 suggests that the plasma flow, denoted by $r_{\rm P}$, can either be greater or smaller than the corotation rate, signified by $r_{\rm C}$, of the star. For plasma flow greater than the corotation rate, $\delta r =r_{\rm P}/r_{\rm C} > 1$ and the plasma flow is ahead of corotation and the subpulses appear to arrive at earlier longitudinal phases. For $\delta r < 1$, the plasma flow lags behind corotation resulting in the movement of the subpulses toward later longitudinal phases. It is difficult to tell whether the changes in the electric drift can be associated with changes in the electric field or in the magnetic field, or both. Electric field in pulsar magnetospheres broadly originates from two models of vacuum and corotation. Furthermore, observations show that the magnetosphere of a pulsar can abruptly change between the two models \citep{Kramer2006}, implying that the electric field can also change. In addition, there are proposals that the magnetic field near the stellar surface is complex and deviates from dipolar structure \citep{Asseo2002, Petri2015}. This may account for the changing direction of the subpulse movement. However, changes in the magnetic field may affect the current flow on a larger scale in the global magnetosphere. Another feature as revealed by the subpulse movement is that the sign change of $\delta r$ appears random. However, during profile shifting, the apparent movement of subpulses is in-phase with the direction of the profile shifting when the emission returning to its original longitudinal phases. The value of $\delta r$ is less than one shortly after a shifting event resulting in the observed subpulses to move toward the leading edge of the profile. This suggests that the profile shifting may be the result of a sudden increase in the flow rate of the subpulses relative to the rotation of the star causing the subpulses to move in and out of the profile window. Identification of the profile shifting in the paper by \citet{CLH+20} was based on deviation in the timing residuals above 2.5 times the weighted RMS. This also implies that the pulse arrival times (ToAs) during profile shifting deviate from average. Zero ToAs would mean that the pulses arrive at the time as predicted, and the presence of timing residuals indicates the difference between predicted and measured pulse arrival times. From Figure 5 in \citet{CLH+20}, the deviation is about 3 milliseconds above the weighted RMS. Interpreting the difference as due to delay (or advance) in the pulse arrival, then this translates to about $0.9^\circ$ for a rotation period of 1.23\,s. This agrees with the average amount of shifting in the profile peak detected in our observations. 

\subsection{Implications for the emission geometry}

It is suggested that the underlying mechanism for drifting subpulses is related to the obliquity angle \citep{Weltevrede2008}. The variation in the sign of $\delta r$ in PSR J0344$-$0901 implies that the direction of the subpulse movement is not likely dependent only on the obliquity angle of a pulsar. The characteristics of pulsar emission are closely related to the physical properties and geometric aspects of the emission beam. The two-lobed appearance in the fluctuation spectra, as opposed to other seemingly complex single profiles, may be an indication of specific physical characteristics or the geometric configuration of the pulsar's emission beam  \citep{RRW06, RWB13}. Notably, pulsars are known to have intricate magnetic field structures that channel their emissions into narrow beams \citep{Asseo2002, Gupta2003}. If our line of sight intersects two distinct regions of the beam, such as an inner core and an outer cone, this could manifest as a two-lobed profile \citep[e.g., B1944+17;] [] {Kloumann10}. The situation may be further compounded by numerous factors, including relativistic beaming effects \citep{Dyks2003}, plasma instabilities in the emission region \citep{Roy2019, Ben2023}, or propagation effects within the pulsar's magnetosphere \citep{Petrova2000}. Consequently, our pulsar provides a unique exemplar of how analyzing emission features can contribute to estimating the radiation structure and its changes within the emission region. Understanding these changes can offer explanations for unconventional emission variations, such as profile shifting and the different subpulse movements observed in PSR J0344$-$0901, thereby providing new information for the interaction between pulsar emission properties and the geometric configurations in the emission beams.

\subsection{Comparison with others radio pulsars}

Recent observations using the FAST telescope have significantly expanded our knowledge of pulsar emission dynamics through discoveries of variations in drifting subpulses and the associated changes in the subpulse drift patterns. For example, PSR J1926$-$0652 \citep{Zhang2019} displays complex subpulse drifting across multiple profile components, with irregular $P_{3}$ and different drift bands, indicating intricate magnetospheric activities. PSR J1631+1252 \citep{Wen2022} and the nulling pulsar PSR B2111+46 \citep{Zhi2023} both demonstrate modulated drifting subpulses in their leading components, revealing the local emission properties. In addition, with the switching in the subpulse drift-rate and curved drift-bands, PSR J1857+0057 \citep{Yan2023} highlights the temporal variation in pulsar emission.

In comparison, PSR J2007+0910 \citep{Xu2024} presents an interesting case with its varying drift-bands, suggesting that diverse emission modes an occur in the same pulsar. This contrasts with the more systematic and predictable curved drift-bands observed in PSR B0809+74 \citep{Hassall2013} and PSR B0826$-$34 \citep{ELG+05}. Furthermore, discoveries by the Arecibo telescope, such as that in PSR B0919+06 and PSR B1859+07, reveal profile shifting towards earlier longitudinal phases \citep{RRW06, Perera15, HHP+16,WOR+16, Shaifullah18}. The phenomenon was suggested in connection with the relativistic aberration and retardation effects in the conventional geometric models \citep{CRZ00, DK04, Gangadhara05}.

In contrast, the unique observations of the sporadic and irregular subpulse movements in PSR J0344$-$0901, which are accompanied with profile shifting, challenge these conventional models. The observation of PSR J0344$-$0901, with emission potentially extending beyond the profile boundary \citep{RRW06}, shows an exceptional aspect of pulsar behavior and a significant difference from previous findings.

Collectively, observations from the FAST telescope, together with the Arecibo findings, reveal a broader range of pulsar behaviors that challenge existing pulsar emission models. Distinct characteristics from each pulsar, especially from the unprecedented observations of PSR J0344$-$0901, contribute to a deeper and more nuanced understanding of pulsar emission processes and the complex dynamics of pulsar magnetospheres.

\section{Summary}
\label{sect:summary}

We have reported on the emission properties in PSR J0344$-$0901 based on single-pulse observations performed at a frequency range centered at 1.25 GHz on two different epochs using FAST.  The pulsar is classified as a normal pulsar, and possesses a surface dipole magnetic field strength of approximately 2.06 $\times$ 10$^{12}$ G. Applying the rotating vector model to the changes in the polarization position angle gave an estimation for the inclination angle of $75.12^\circ \pm 3.80^\circ$ with an impact parameter of  $-3.17^\circ \pm 5.32^\circ$ between the line-of-sight and the magnetic axis. This agrees with the suggestion that the inclination angle of a pulsar is correlated with the strength of its magnetic field \citep{Zhang1998}. In addition, the pulsar exhibits significant variation in single-pulse emission, characterized by profile shifting and different subpulse movement patterns.  Key emission characteristics of the pulsar include: 
	
	\begin{itemize}[leftmargin=0.5cm]
		
		\item Rotation period of 1.23\,s.
		\item Pulse nulling was not detected in the pulsar.
		
		\item Shifting of profile to later longitudinal phases was detected, with an average shift of about 0.7$^\circ$ measured at the profile peak from the original position.
		\item On average, the profile shifting lasts for about 216 pulse periods or approximately 265 seconds.
		\item The occurrence frequency and the duration for different events of profile shifting vary significantly, with shifting emission accounting for about 38.78\% of observation time. 
		
		\item Noticeable changes in the polarization position angle were detected during profile shifting.
		
		\item  We identify four distinct emission modes, referred to as Modes A, B, C, and D, each with unique subpulse movement feature and different subpulse track pattern. Mode D is the most abundant mode in our observations.
		
		\item  Subpulses in Modes A and B exhibit movement toward later and earlier longitudinal phases, respectively. Mode C features consistently curved subpulses tracks, while Mode D displays irregular subpulse tracks.
		
	\end{itemize}

	\section*{Acknowledgments}

	This work is supported by the National Natural Science Foundation of China (Grant Nos. 11988101, U2031117, U1838109, 11873080, 12041301) and by the Alliance of International Science Organizations, Grant No. ANSO-VF-2024-01. We are grateful to the anonymous referee for valuable comments that have improved the presentation of this paper. H.M.T. acknowledges Arba Minch University, the University of Chinese Academy of Sciences, and the CAS-TWAS President's Fellowship Programme for providing funding and a PhD Scholarship (No. 2019A8016609001). R.Y. is supported by the National SKA Program of China (No. 2020SKA0120200), the National Key Program for Science and Technology Research and Development (No. 2022YFC2205201), the National Natural Science Foundation of China (NSFC) project (Nos. 12041303, 12288102), the Major Science and Technology Program of Xinjiang Uygur Autonomous Region (No. 2022A03013-2), the Open Program of the Key Laboratory of Xinjiang Uygur Autonomous Region (No. 2020D04049), the Major Science and Technology Program of Xinjiang Uygur Autonomous Region No. 2022A03013-2, and is partly supported by the Operation, Maintenance, and Upgrading Fund for Astronomical Telescopes and Facility Instruments, which is budgeted from the Ministry of Finance of China (MOF) and administered by the CAS. N.W. is supported by the National Natural Science Foundation of China (NSFC Grant No. 12041304). D.L. is supported by the National Natural Science Foundation of China (NSFC Grant No. 11988101) and the 2020 project of Xinjiang Uygur Autonomous Region of China for flexibly fetching in upscale talents. P. Wang is supported by the National SKA Program of China (Grant No. 2020SKA0120200), the Youth Innovation Promotion Association CAS (id. 2021055), and the CAS Project for Young Scientists in Basic Research (grant YSBR-006). We thank the XAO pulsar group for valuable discussions. We also express our gratitude to all members of the FAST telescope collaboration for establishing the projects (project numbers: ZD2020-06 and PT2020-0045), which made the observations possible.

	This work made use of data from the Five-hundred-meter Aperture Spherical radio Telescope (FAST). FAST is a Chinese national mega-science facility operated by the National Astronomical Observatories, Chinese Academy of Sciences.

	\bibliographystyle{aasjournal}
	\bibliography{Profile_shifting}
	\end{document}